\documentclass[twocolumn,aps,prc,superscriptaddress,showpacs,floatfix,longbibliography,nofootinbib]{revtex4-1}
\usepackage{url}
\usepackage{cancel}
\usepackage[colorlinks,linkcolor=blue,citecolor=blue,filecolor=black,urlcolor=blue]{hyperref}
\usepackage{epsfig,graphics}
\usepackage{graphicx}
\usepackage{dcolumn}
\usepackage{bm}
\usepackage[usenames]{color}
\usepackage{amssymb}
\usepackage{amsmath}
\usepackage{multirow}
\usepackage{float}
\usepackage{harpoon}
\usepackage{MnSymbol}
\usepackage{appendix}
\usepackage{color}
\usepackage{hyperref}
\usepackage{cleveref}
\usepackage{dutchcal}

\begin{document}

\title{Spin polarization from nucleon-nucleon scatterings in intermediate-energy heavy-ion collisions}
\author{Rong-Jun Liu}
\affiliation{Shanghai Institute of Applied Physics, Chinese Academy of Sciences, Shanghai 201800, China}
\affiliation{University of Chinese Academy of Sciences, Beijing 100049, China}
\author{Jun Xu}\email[Correspond to\ ]{junxu@tongji.edu.cn}
\affiliation{School of Physics Science and Engineering, Tongji University, Shanghai 200092, China}
\affiliation{Shanghai Institute of Applied Physics, Chinese Academy of Sciences, Shanghai 201800, China}
\author{Yu-Gang Ma}
\affiliation{Key Laboratory of Nuclear Physics and Ion-beam Application (MOE), Fudan University, Shanghai 200433, China}
\affiliation{Shanghai Research Center for Theoretical Nuclear Physics, NSFC and Fudan University, Shanghai 200438, China}
\begin{abstract}
We propose a new mechanism of generating spin polarization in heavy-ion collisions dominated by nucleon degree of freedom. By incorporating the spin change in nucleon-nucleon scatterings based on the phase shift data together with the constraint of rigorous angular momentum conservation and Pauli blocking, we illustrate through a Boltzmann-Uehling-Uhlenbeck transport model that appreciable spin polarization (about $1 \sim 2\%$) can be generated in intermediate-energy heavy-ion collisions. This mechanism, together with the nuclear spin-orbit potential, may help to understand the spin polarization in few-GeV heavy-ion collisions dominated by nucleon degree of freedom.
\end{abstract}
\maketitle


Spin physics has been an interesting topic in various research fields. In recent years, the spin polarization phenomena in relativistic heavy-ion collisions have attracted considerable attentions~\cite{Becattini:2020ngo}. The large angular momentum in non-central relativistic heavy-ion collisions generates a vorticity field in the produced quark-gluon plasma~\cite{Jiang:2016woz}, leading to the global spin polarization of the participant matter perpendicular to the reaction plane. Such spin polarization has observational consequences for particles produced from the hadronization of the partonic phase. While the spin states of hyperons or vector mesons can be affected by the hadronic evolution~\cite{PhysRevC.95.054902,PhysRevC.100.014913,Becattini:2019ntv,Sung:2024vyc}, their appreciable spin polarizations or spin alignments were observed experimentally through the angular distribution of their decays~\cite{STAR:2017ckg,STAR:2022fan,PhysRevLett.126.162301,PhysRevC.98.014910,ALICE:2019aid,STAR:2023nvo} (see also theoretical studies in Refs.~\cite{PhysRevLett.94.102301,PhysRevLett.117.192301,PhysRevC.96.024906,Sheng:2022wsy,Chen:2023,Chen:2024afy}). More recently, it is interesting to see that even stronger spin polarizations of hyperons were observed in heavy-ion collisions at lower energies~\cite{PhysRevC.104.L061901,2022137506} dominated by nucleon degree of freedom, likely due to the stronger stopping and vorticity~\cite{Deng:2021miw,Deng:2020ygd}.

Knowledge of the spin dynamics in heavy-ion collisions dominated by nucleons is essential in understanding both the effect of the hadronic afterburner on the particle spin in relativistic heavy-ion collisions and the spin-related observables at lower collision energies. The nuclear spin-orbit potential is one of the major sources of the spin polarization in intermediate-energy heavy-ion collisions~\cite{Xia:2019whr,Xu:2025uwd}, where the effect of the magnetic field is expected to be small. In this manuscript, we present a new mechanism that may lead to considerable nucleon spin polarizations. Starting from the phase shift data, we find that appreciable nucleon spin polarizations can be generated through nucleon-nucleon (NN) scatterings in intermediate-energy heavy-ion collisions even in the absence of the nuclear spin-orbit potential, if the rigorous angular momentum conservation (AMC) and the Pauli blocking are properly incorporated.


We start from the spin state of a nucleon expressed as
\begin{equation}
\left| \psi \right> =\exp \left( -i\phi_s /2 \right) \cos \left( \theta_s /2 \right) \left| z,+ \right> +\exp \left( i\phi_s /2 \right) \sin \left( \theta_s /2 \right) \left| z,- \right>,
\end{equation}
where the $z$ direction is the spin projection direction in the heavy-ion collision system, and $\phi_s$ and $\theta_s$ are respectively the azimuthal and polar angles representing the spin expectation direction. The spin density matrix can then be expressed as
\begin{equation}
\rho ^s=\left| \psi \right> \left< \psi \right|.
\end{equation}
The above spin density matrix can be transformed to the helicity density matrix through the relation
\begin{equation}
\rho ^h=D\rho ^sD^+,
\end{equation}
where $D$ is the rotation matrix which changes the $z$ axis to the momentum direction $\hat{p}$ of the nucleon in the center-of-mass (c.m.) frame of the NN scattering
\begin{equation}
D=\left( \begin{matrix}
	\exp \left( i\phi _p/2 \right) \cos \left( \theta _p/2 \right)&		\exp \left( -i\phi _p/2 \right) \sin \left( \theta _p/2 \right)\\
	-\exp \left( i\phi _p/2 \right) \sin \left( \theta _p/2 \right)&		\exp \left( -i\phi _p/2 \right) \cos \left( \theta _p/2 \right)\\
\end{matrix} \right),
\end{equation}
with $\phi_p$ and $\theta_p$ being the azimuthal and polar angles of $\hat{p}$ before the NN scattering. We note that the way of rotating the coordinate system is not unique, e.q., further rotations of the coordinate system around $\hat{p}$ are allowed, and they may lead to different helicity density matrices, but do not change the final spin expectation direction after the NN scattering. The total helicity density matrix of the NN scattering can be obtained by taking the direct product, i.e.,
\begin{equation}
\rho ^h\left( A,B \right) =\rho ^h\left( A \right) \otimes \rho ^h\left( B \right),
\end{equation}
and the final-state helicity density matrix in the $A+B \rightarrow C+D$ process can be expressed as
\begin{equation}
\rho ^h\left( C,D \right)=\frac{{H\rho ^h\left( A,B \right)H^{*}}}{tr\left[ {H\rho ^h\left( A,B \right)H^{*}} \right]},
\end{equation}
where
\begin{eqnarray}
&&H_{\lambda _C\lambda _D;\lambda _A\lambda _B}\left( \theta_p^\prime ,\phi_p^\prime \right) \notag\\
&=&\exp \left[ i\left( s_B-s_D \right) \right] \sqrt{\frac{\pi}{pp'}}\frac{e^{i\phi_p^\prime (\lambda-\mu)}}{p} \notag\\
&\times& \sum_{J=0}^{\infty}{\left( 2J+1 \right) \left< JM\lambda _C\lambda _D \right|T\left| JM\lambda _A\lambda _B \right> d_{\lambda \mu}^{J}\left( \theta_p^\prime \right)}
\end{eqnarray}
is the helicity amplitude~\cite{Leader:2011vwq}. 
In the above, $\phi_p^\prime$ and $\theta_p^\prime$ are the azimuthal and polar angles of the nucleon momentum direction after the NN scattering, $s_B$ and $s_D$ are respectively the spins of nucleon $B$ and $D$, $p$ and $p'$ are respectively the momentum magnitudes in the c.m. frame before and after NN scattering, $\lambda =\lambda_A-\lambda_B$ and $\mu =\lambda_C-\lambda _D$ are calculated from $\lambda_{A,B,C,D}$ which are the helicities and the indices of the matrix element, $d_{\lambda \mu}^{J}$ is the Wigner rotation matrix, and $J$ and $M$ are respectively the angular momentum and magnetic quantum numbers. The quantum number $M$ in the expression will not participate in the final result, as mentioned in Ref.~\cite{Jacob:1959at}. The relationship between the T matrix and the S matrix is $S-1=iT$, and the S matrix can be expressed as
\begin{eqnarray}\label{Smatrix}
&&\left< JM\lambda _C\lambda _D \right|S\left| JM\lambda _A\lambda _B \right> \notag\\
&=&\sum_{J_1,M_1,L_1,S_1,J_2,M_2,L_2,S_2}\left< JM\lambda _C\lambda _D\mid J_2M_2L_2S_2 \right> \notag\\
&\times& \left< J_2M_2L_2S_2 \right|S\left| J_1M_1L_1S_1 \right> \left< J_1M_1L_1S_1\mid JM\lambda _A\lambda _B \right>, \notag\\
&=&\sum_{L_1,S_1,L_2,S_2}\sqrt{\frac{2L_2+1}{2J+1}}\sqrt{\frac{2L_1+1}{2J+1}}\left( L_10S_1\lambda |J\lambda \right) \left( L_20S_2\mu |J\mu \right) \notag\\
&\times&\left( s_A\lambda _As_B-\lambda _B|S_1\lambda \right) \left( s_C\lambda _Cs_D-\lambda _D|S_2\mu \right) S_{L_2S_2;L_1S_1}^{J},
\end{eqnarray}
where we have used the relation~\cite{chung2008spin} 
\begin{eqnarray}
&&\left< J_1M_1L_1S_1 \mid JM\lambda _A\lambda _B \right> \notag\\
&=&\sqrt{\frac{2L_1+1}{2J+1}}\left( L_10S_1\lambda |J\lambda \right) \left( s_A\lambda _As_B-\lambda _B|S_1\lambda \right) \delta _{JJ_1}\delta _{MM_1}. \notag\\
\end{eqnarray}
In the above, the parentheses represent the Clebsch-Gordan coefficient, and $\left< JML_2S_2 \right|S\left| JML_1S_1 \right> $ has been expressed as $S_{L_2S_2;L_1S_1}^{J}$, where $M$ has no contribution. In this way, we transform the S matrix for the helicity spin state to that for the canonical spin state, and the latter can be expressed in terms of the phase shift $\delta$. For the spin-singlet or spin-triplet case with $L=J$, there is only one channel $S=\exp(2i\delta)$. For the spin-triplet case with $L=J\pm1$, there are two channels~\cite{Xia:2017dbx}
\begin{widetext}
\begin{equation}
S=\left( \begin{matrix}
	\cos ^2(\epsilon) e^{2i\delta _{J-1}}+\sin ^2(\epsilon) e^{2i\delta _{J+1}}&		\frac{1}{2}\sin \left( 2\epsilon \right) \left( e^{2i\delta _{J-1}}-e^{2i\delta _{J+1}} \right)\\
	\frac{1}{2}\sin \left( 2\epsilon \right) \left( e^{2i\delta _{J-1}}-e^{2i\delta _{J+1}} \right)&		\sin ^2(\epsilon) e^{2i\delta _{J-1}}+\cos ^2(\epsilon) e^{2i\delta _{J+1}}\\
\end{matrix} \right),
\end{equation}
\end{widetext}
where $\epsilon$ is the mixing factor describing the mixing probability of the two coupling states. We adopt the phase-shift data and mixing factor from proton-neutron scatterings ranging from the energy 0 to 425 MeV, and proton-proton scatterings ranging from the energy 1 to 500 MeV~\cite{PhysRevC.15.1002}, with the latter assumed to be the same for neutron-neutron scatterings. In this way, we get the helicity density matrix $\rho^h(C,D)$ of the final state. We can use the previous rotation matrix $D$ to transform it back into a spin density matrix, and then use the Pauli matrix to get the spin expectation direction of the final-state nucleon
\begin{eqnarray}
\vec{s}^\star_C&=&tr\left( \left( \vec{\sigma}\otimes \hat{I} \right) D_B^+ \otimes D_A^+ \rho^h(C,D)D_A \otimes D_B \right),\label{sc}\\
\vec{s}^\star_D&=&tr\left( \left( \hat{I}\otimes \vec{\sigma} \right) D_B^+ \otimes D_A^+ \rho^h(C,D)D_A \otimes D_B \right),\label{sd}
\end{eqnarray}
where $\hat{I}$ is the identity matrix, and $\vec{\sigma}$ represents the Pauli matrices. Note that the magnitude of $\vec{s}^\star_{C,D}$ may not be 1 and should be normalized as $\vec{s}_{C,D} = \vec{s}^\star_{C,D}/|\vec{s}^\star_{C,D}|$. We adopt the above framework only for elastic NN scatterings in the present study.

\begin{figure}[ht]
\includegraphics[width=0.8\linewidth]{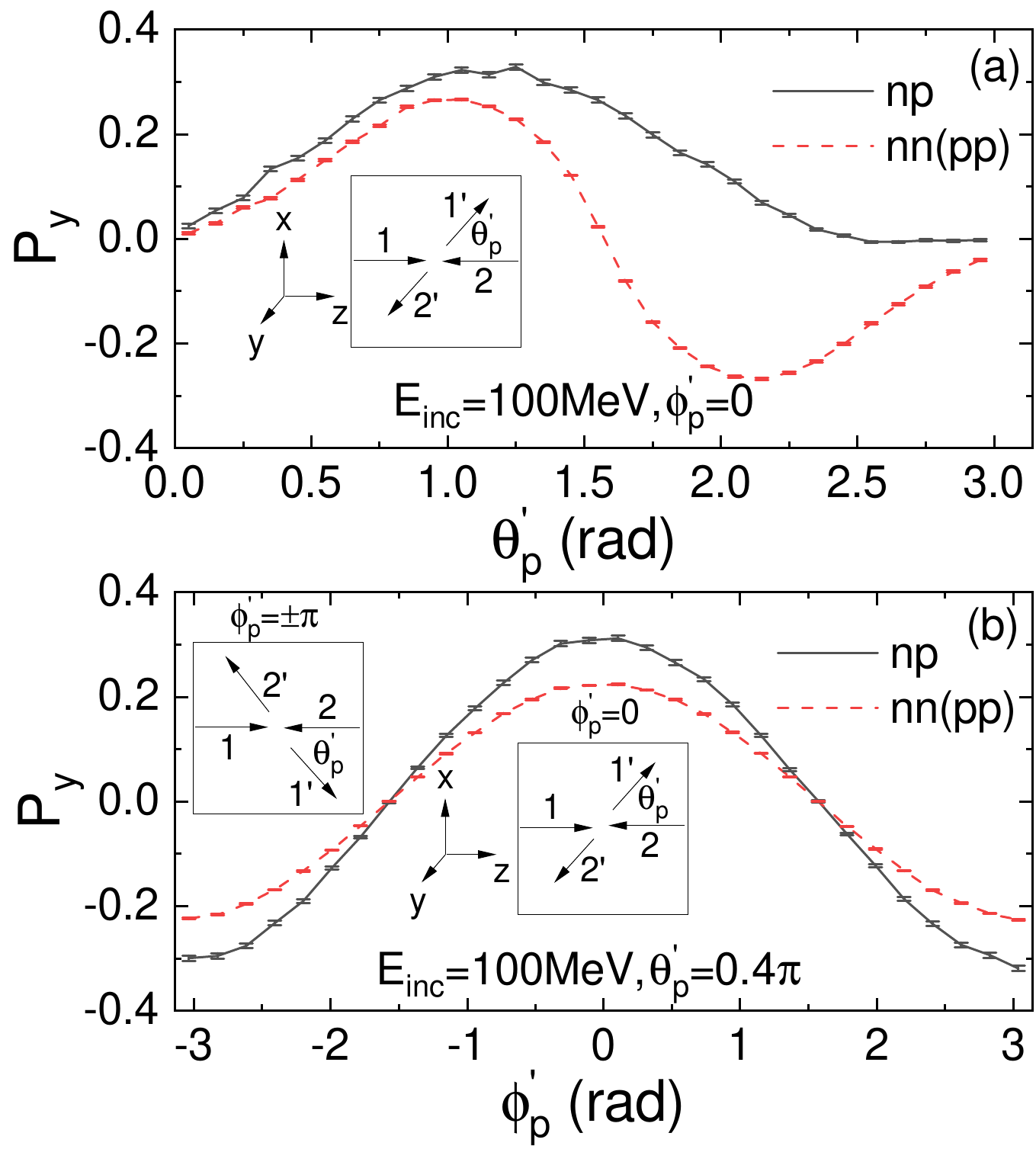}
\caption{\label{NN} Polarization in $y$ direction after neutron-proton (np) or neutron-neutron (nn) (proton-proton (pp)) scatterings: (a) Polar angular dependence at the nucleon incident energy of 100 MeV and $\phi_p^\prime=0$; (b) azimuthal angular dependence at the nucleon incident energy of 100 MeV and $\theta_p^\prime=0.4\pi$.}
\end{figure}

\begin{figure}[ht]
\includegraphics[width=0.8\linewidth]{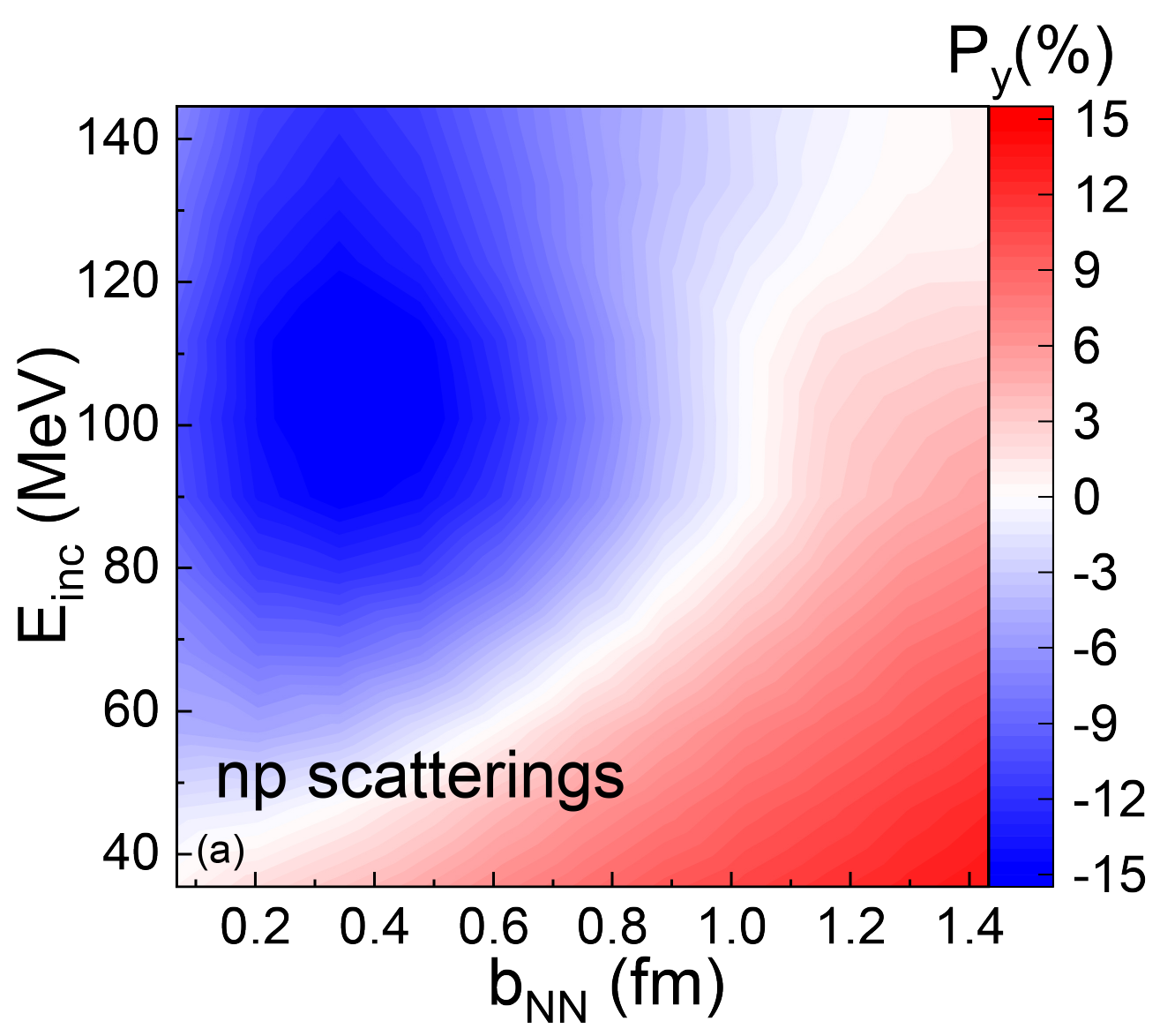}\\
\includegraphics[width=0.8\linewidth]{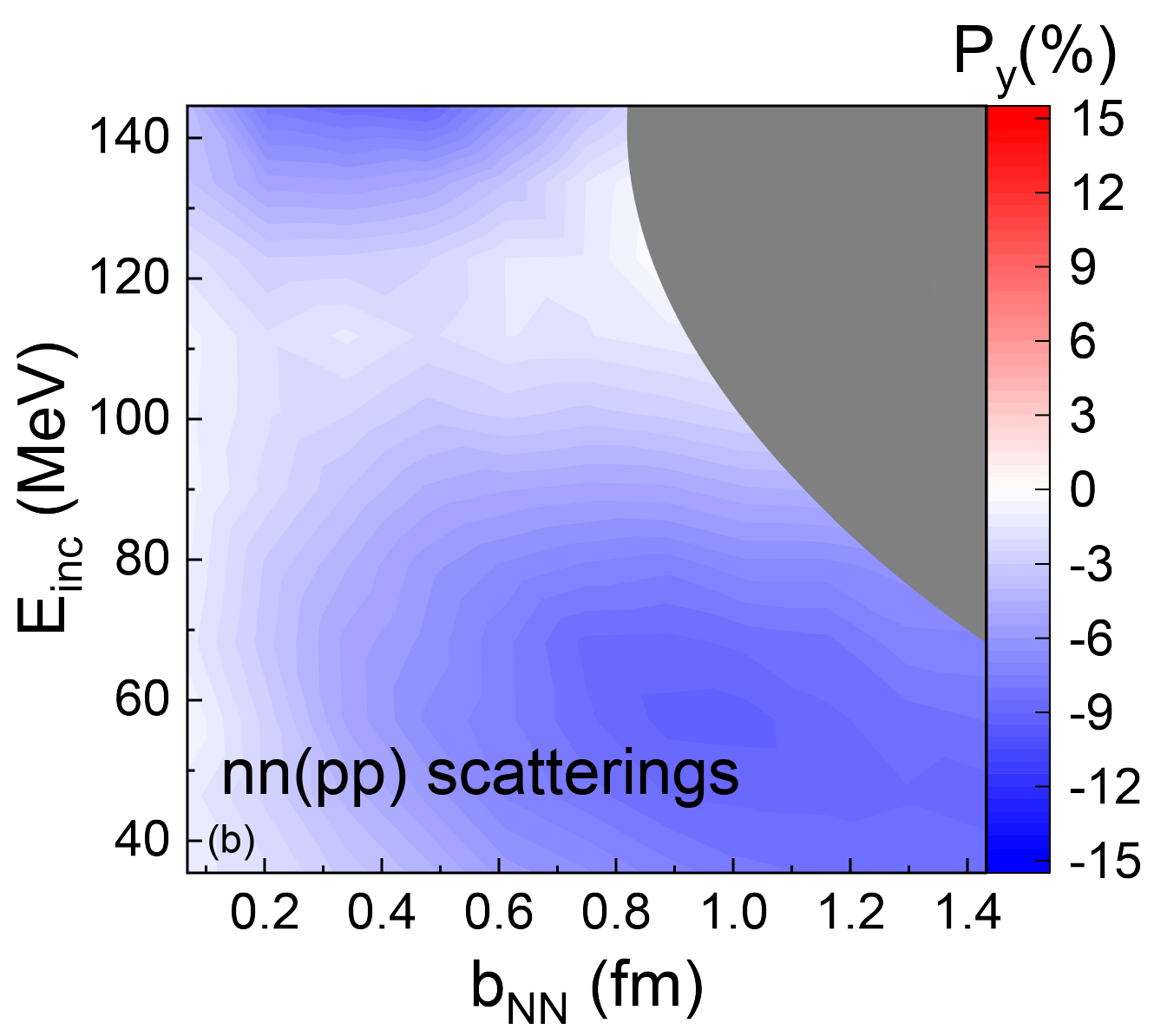}
\caption{\label{NN_AMC} Net spin polarization in $y$ direction after np (a) and nn (pp) (b) scatterings at different NN impact parameters $\text{b}_{NN}$ and incident energies $E_{inc}$, after incorporating the constraint of AMC and integrating all possible initial nucleon spins and final nucleon momentum directions.}
\end{figure}

Figure~\ref{NN} illustrates the spin polarization $P_y$ of final-state nucleons in $y$ direction by using 0.5 million NN scatterings based on the above framework, where the error bars represent the uncertainty due to the random spin directions of nucleons before NN scatterings. In the c.m. frame of the NN scattering, the $z$ direction is set as the initial momentum direction of the first nucleon, and the polar angle $\theta_p^\prime$ and azimuthal angle $\phi_p^\prime$ represent its final-state momentum direction $\hat{p'}$, with those of the second nucleon being in the opposite direction. Taking the incident energy $E_{inc}=100$ MeV as an example, it is seen that for the in-plane ($\phi_p^\prime=0$) case neutron-proton (np) scatterings leads to an overall positive $P_y$ which peaks around $\theta_p^\prime=0.4\pi$, while neutron-neutron (nn) or proton-proton (pp) scatterings lead to an overall zero $P_y$ with positive (negative) values at small (large) $\theta_p^\prime$, as shown in Fig.~\ref{NN} (a). Thus, it is expected that np scatterings may lead to considerable spin polarizations while nn or pp scatterings have much smaller contributions. For a fixed $\theta_p^\prime=0.4\pi$, it is seen that $P_y$ has a peak around $\phi_p^\prime=0$ and a valley around $\phi_p^\prime=\pm\pi$.

The rigorous AMC without spin requires in-plane ($\phi_p^\prime=0$ or $\pm\pi$) NN scatterings~\cite{Gale:1990zz}, and incorporating the nucleon spin degree of freedom may lead to only minor deviations~\cite{PhysRevC.109.014615}. Thus, while the integrated $P_y$ is zero in Fig.~\ref{NN} (b), incorporating the rigorous AMC only allows final momentum directions around $\phi_p^\prime=0$ and $\phi_p^\prime=\pm\pi$, and may affect the polar angular distribution of NN scatterings as well. This may lead to non-zero $P_y$, i.e., spin polarization in the direction of $\pm \hat{p} \times \hat{p'}$ according to the cartoon shown in Fig.~\ref{NN} (b). We use the spin-dependent differential neutron-neutron (proton-proton) and neutron-proton elastic scattering cross sections, with the polar angular dependence as parametrized in Ref.~\cite{Xia:2017dbx} from the same phase shift data~\cite{PhysRevC.15.1002}. Accordingly, we first sample the nucleon momenta after NN scatterings as in Ref.~\cite{Bertsch:1988ik}, and the final-state nucleon spins can then be obtained using Eqs.~(\ref{sc}) and (\ref{sd}). To incorporate the constraint of AMC~\cite{PhysRevC.109.014615}, the coordinates and momenta of final-state nucleons after NN scatterings need to be slightly adjusted with given final-state nucleon spins. Thus, the nucleon momenta and spins after NN scatterings should be self-consistently obtained through iteration, and a good convergence can be achieved after about 10 iterations. Integrating all possible initial spins and final momentum directions of scattering nucleons, the net spin polarizations in $y$ direction after np and nn (pp) scatterings in the plane of NN impact parameter $\text{b}_{NN}$ and incident energy $E_{inc}$ after incorporating the constraint of AMC are displayed in Fig.~\ref{NN_AMC}. The shaded area means that the cross section is not large enough to allow scatterings to happen, so $P_y$ is meaningless in the area. While $P_y$ in most region is negative, in neutron-rich matter np scatterings dominate the overall scatterings, and there are more scatterings with larger NN impact parameters and lower scattering energies in real heavy-ion collisions, where one expects that the region with small $E_{inc}$ and large $\text{b}_{NN}$ for np scatterings may lead to a positive $P_y$.


We now move to the illustration of the effect on nucleon spin polarization in intermediate-energy heavy-ion collisions, for which the simulation is based on the Boltzmann-Uehling-Uhlenbeck equation~\cite{TMEP:2022xjg,Xu:2019hqg}
\begin{equation}
  \label{eq:boltz}
  \frac{\partial f_\tau(\vec{r},\vec{p})}{\partial t}
  +\frac{\vec{p}}{\sqrt{m^2+\vec{p}^2}}\cdot
  \frac{\partial f_\tau(\vec{r},\vec{p})}{\partial \vec{r}}-\frac{\partial U_\tau}{\partial \vec{r}}\cdot\frac{\partial f_\tau(\vec{r},\vec{p})}{\partial \vec{p}}
  =I_c.
\end{equation}
The left-hand side describes how the phase-space distribution function $f_\tau(\vec{r},\vec{p})$ for nucleons with bare mass $m$ and isospin index $\tau$ ($\tau=n,p$) evolves with time under the mean-field potential $U_\tau$. Here we adopt the spin-independent mean-field potential in the nuclear matter of density $\rho$ and isospin asymmetry $\delta$ as
\begin{equation}\label{u}
U_{n,p}(\rho,\delta)=\alpha \left(\frac{\rho}{\rho_{0}}\right)+\beta\left(\frac{\rho}{\rho_{0}}\right)^{\gamma} \pm 2 E_{sym}^{pot}  \left(\frac{\rho}{\rho_{0}}\right)^{\gamma_{sym}} \delta, 
\end{equation}
with the `+' (`$-$') sign for neutrons (protons), and $\alpha=-209.2$ MeV, $\beta=156.4$ MeV, $\gamma=1.35$, $E_{sym}^{pot}=18$ MeV, and $\gamma_{sym}=2/3$ reproducing the empirical nuclear matter properties. For the $I_c$ part, we incorporate elastic NN scatterings with the differential cross sections and spin changes as described above, with Pauli blocking properly incorporated. In the initialization of heavy-ion simulations, the nucleon coordinates are sampled according the density distribution generated by the Skyrme-Hartree-Fock calculation~\cite{Chen:2010qx}, and the nucleon momenta are sampled within the local Fermi sphere. The initial nucleon spin expectation directions are randomly sampled in the $4\pi$ solid angle.

\begin{figure}[ht]
\includegraphics[width=0.8\linewidth]{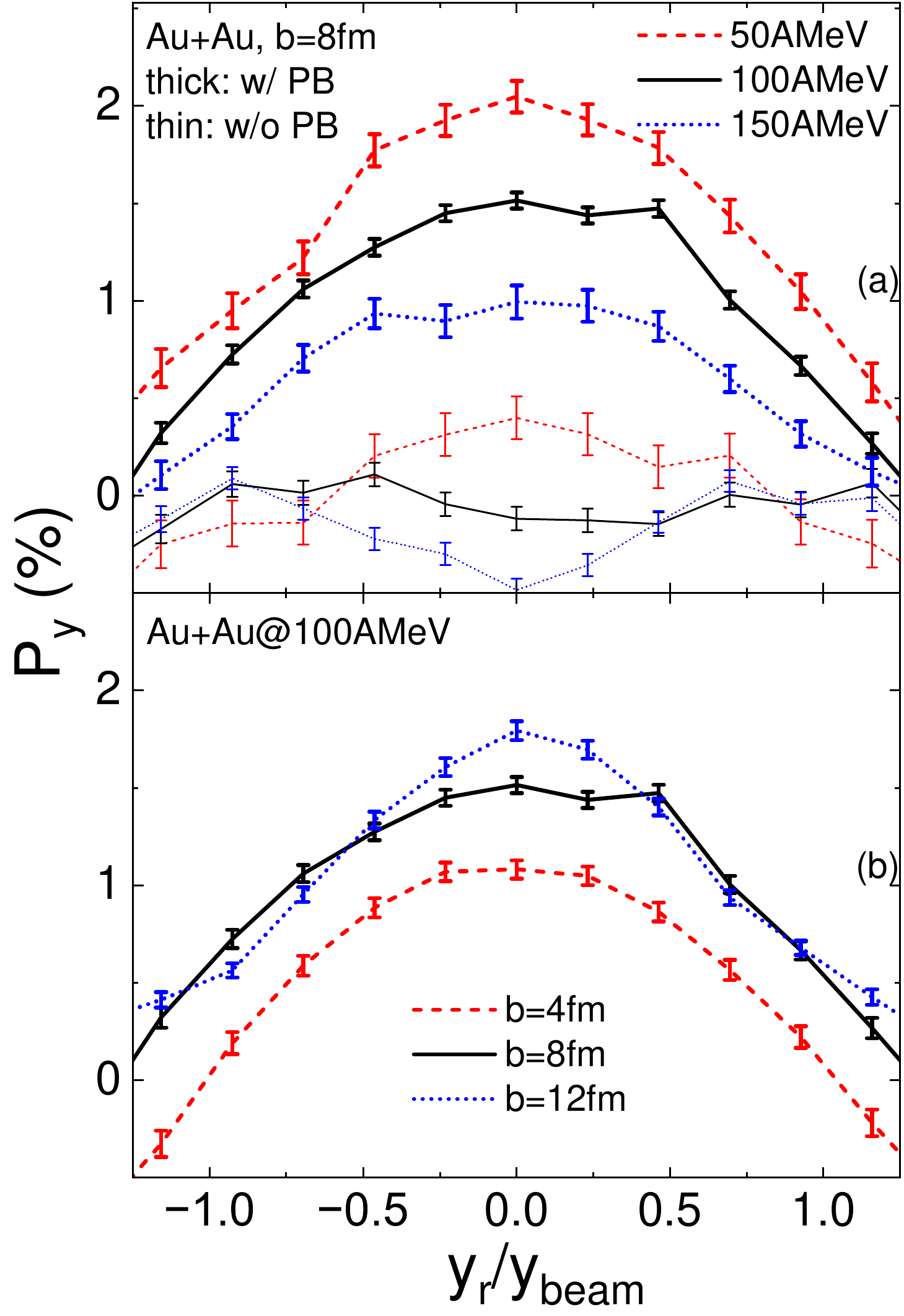}
\caption{\label{Py-y} Spin polarization of free nucleons perpendicular to the reaction plane as a function of reduced rapidity in Au+Au collisions. Upper: At different beam energies and a fixed impact parameter $\text{b} = 8$ fm with and without Pauli blocking (PB); Lower: At different impact parameters and a fixed beam energy 100 AMeV with Pauli blocking.}
\end{figure}

Figure~\ref{Py-y} compares the spin polarization perpendicular to the reaction plane for free nucleons, the densities of which are below $1/8$ of the saturation density in the final state of heavy-ion collisions. In the default setup with Pauli blocking, considerable spin polarization $P_y$ is observed at midrapidities. This can be understood as the large orbital angular momentum in heavy-ion collisions transfers to the nucleon spin polarization through NN scatterings. Since the nucleon spin tends to be polarized in the direction of $\hat{p}\times\hat{p'}$, NN scatterings with both $\hat{p}$ and $\hat{p'}$ in the reaction plane contribute significantly to the total $P_y$. $P_y$ is larger at midrapidities since it is produced from NN scatterings in the participant region, while in the large-rapidity spectator region $P_y$ is largely reduced. $P_y$ increases with decreasing beam energy as a result of the larger NN cross section and thus larger NN impact parameter $\text{b}_{NN}$ as well as the stronger Pauli blocking at lower beam energies. At larger impact parameters of heavy-ion collisions, $P_y$ at midrapidities becomes even larger, since the effect of nucleon Fermi motion, which smears the orbital angular momentum in NN scatterings, becomes weaker. If the Pauli blocking is turned off, $P_y$ is negligibly small, and the reason will be discussed in the following.

\begin{figure}[ht]
\includegraphics[width=0.8\linewidth]{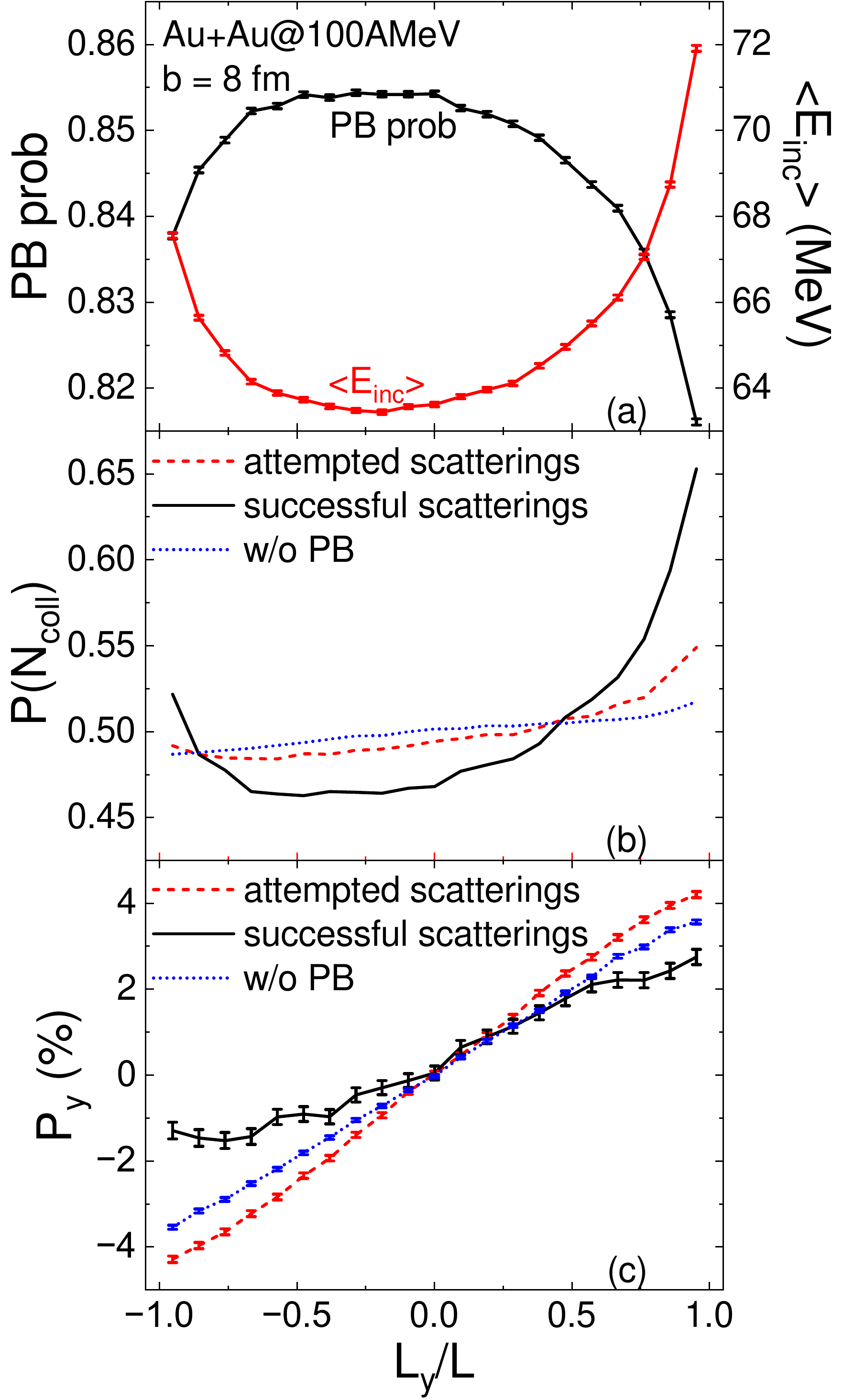}
\caption{\label{pauli} Upper: Dependence of the Pauli blocking probability (PB prob) and the average incident energy of colliding nucleons on the y-component fractional angular momentum $L_y/L$ of NN scatterings. Middle: Dependence of normalized attempted and successful NN scattering numbers as well as the normalized NN scattering number in the absence of Pauli blocking (w/o PB) on $L_y/L$ of NN scatterings. Bottom: Dependence of the final-state spin polarization perpendicular to the reaction plane on $L_y/L$ of NN scatterings. Results are from IBUU simulations for Au+Au collisions at the beam energy of 100 AMeV and impact parameter $\text{b}=8$ fm.}
\end{figure}

Taking midcentral Au+Au collisions at the beam energy of 100 AMeV as an example, Fig.~\ref{pauli} (a) shows that the Pauli blocking probability is different for NN scatterings with different y-component fractional angular momentum $L_y/L$ of NN scatterings. While NN scatterings with $L_y/L \sim 0$ have the largest Pauli blocking probability, those with $L_y/L \sim 1$ generally have a smaller Pauli blocking probability than those with $L_y/L \sim -1$. The Pauli blocking probability is strongly anticorrelated with the incident energy $E_{inc}$, since the final state of energetic nucleon scatterings generally have a larger phase space and these scatterings are less Pauli blocked. The larger $\langle E_{inc} \rangle$ for $L_y/L \sim 1$ than that for $L_y/L \sim -1$ is intuitively understandable, since in the former case nucleons move along with the vorticity in the participant region of heavy-ion collisions and are thus more energetic. The difference in the normalized NN scattering numbers $P(N_{coll})$ around $L_y/L \sim 1$ and $L_y/L \sim -1$ is very small for attempted scatterings or in the case without Pauli blocking, while Pauli blocking leads to a considerable difference as shown in Fig.~\ref{pauli} (b). As shown in Fig.~\ref{NN_AMC} (a), the spin polarization in the direction of the orbital angular momentum of NN scatterings mostly originates from low-energy np scatterings. Therefore, NN scatterings with $L_y/L \sim 1$ lead to a net positive $P_y$ and those with $L_y/L \sim -1$ lead to a net negative $P_y$, as shown in Fig.~\ref{pauli} (c). In the presence of Pauli blocking, the large difference in $P(N_{coll})$ at $L_y/L \sim 1$ and $L_y/L \sim -1$ together with the positive slope of $P_y \sim L_y/L$ leads to the appreciable spin polarization at midrapidities as shown in Fig.~\ref{Py-y}. Although $P_y \sim L_y/L$ has a larger slope for total attempted scatterings or for that without Pauli blocking, the spin polarizations of free nucleons are negligibly small as shown in Fig.~\ref{Py-y}, since $P(N_{coll})$ are very similar around $L_y/L \sim 1$ and $L_y/L \sim -1$ as shown in Fig.~\ref{pauli} (b).


To summarize, we have incorporated the spin change in NN scatterings based on the phase shift data for np and nn (pp) scatterings, and have illustrated that this may lead to about $1-2\%$ spin polarization perpendicular to the reaction plane for free nucleons in intermediate-energy heavy-ion collisions, as long as the rigorous angular momentum conservation and Pauli blocking are properly incorporated. The present study has proposed a more realistic treatment of spin change in NN scatterings compared to the previous work~\cite{PhysRevC.109.014615}, and the magnitude of the nucleon spin polarization is smaller but comparable to that with only nuclear spin-orbit potential~\cite{Xu:2025uwd}. Studies with in-medium phase shifts are in progress, but qualitative results are expected to remain similar. Our study provides a new mechanism of generating spin polarization in heavy-ion collisions dominated by nucleon degree of freedom. This mechanism, together with the nuclear spin-orbit potential, could be helpful for understanding the measurable spin polarizations of $\Lambda$ hyperons~\cite{PhysRevC.104.L061901,2022137506} and hypertritons~\cite{Sun:2025oib} in few-GeV heavy-ion collisions.

This work is supported by the National Key Research and Development Program of China under Grants No. 2023YFA1606701, the National Natural Science Foundation of China under Grant Nos. 12375125, 12147101, 11922514, and 11475243, the Guangdong Major Project of Basic and Applied Basic Research No. 2020B0301030008, and the Fundamental Research Funds for the Central Universities.

\bibliography{sibuu_NN}
\end{document}